\documentclass[11pt,a4paper]{article}
\pdfoutput=1
\usepackage{amsmath,amssymb}
\usepackage{graphicx,color}
\usepackage{verbatim}
\usepackage[bf]{caption}

\usepackage{epsfig}

\usepackage[breaklinks,bookmarks=true,colorlinks=true,linkcolor=blue,citecolor=blue,urlcolor=blue,hyperfigures=true]{hyperref}
\usepackage[nosort]{cite} %
\usepackage[bulletsep]{collref} 

\numberwithin{equation}{section}

\ifx\href\asklfhas\newcommand{\href}[2]{#2}\fi

\definecolor{pink}{rgb}{0.7,0,0.7}
\definecolor{green}{rgb}{0,0.5,0}
\definecolor{orange}{rgb}{1,0.4,0.3}

\newcommand{\B}{\color{black}}

\newcommand{\be}{\begin{equation}}
\newcommand{\ee}{\end{equation}}
\newcommand{\ba}{\begin{aligned}}
\newcommand{\ea}{\end{aligned}}
\newcommand{\ben}{\begin{displaymath}}
\newcommand{\een}{\end{displaymath}}
\newcommand{\bea}{\begin{eqnarray}}
\newcommand{\eea}{\end{eqnarray}}
\newcommand{\bean}{\begin{eqnarray*}}
\newcommand{\eean}{\end{eqnarray*}}
\newcommand{\bpmat}{\begin{pmatrix}}
\newcommand{\epmat}{\end{pmatrix}}

\newcommand{\f}{\frac}
\newcommand{\p}{\partial}

\newcommand{\AdS}{{\ensuremath{\text{AdS}}}}
\newcommand{\Sph}{{\ensuremath{\text{S}}}}

\newcommand{\AdSxS}{\ensuremath{{\rm AdS} \times {\rm S}}}
\newcommand{\AdSfiveS}{\ensuremath{{\rm AdS}_5 \times {\rm S}^5}}
\newcommand{\AdSthreeS}{\ensuremath{{\rm AdS}_3 \times {\rm S}^3}}

\newcommand{\SU}{{\ensuremath{\text{SU}}}}
\newcommand{\SO}{{\ensuremath{\text{SO}}}}

\newcommand{\OSP}{{\ensuremath{\text{OSP}}}}

\newcommand{\psu}{{\ensuremath{\mathfrak{psu}}}}

\newcommand{\osp}{{\ensuremath{\mathfrak{osp}}}}

\newcommand{\calN}{{\ensuremath{\mathcal{N}}}}

\renewcommand{\l}{\lambda}
\renewcommand{\th}{\theta}
\renewcommand{\a}{\alpha}

\renewcommand{\b}{\beta}

\newcommand{\g}{\gamma}

\newcommand{\s}{\sigma}

\newcommand{\Th}{\Theta}
\newcommand{\m}{\mu}

\renewcommand{\k}{\kappa}

\long\def\symbolfootnote[#1]#2{\begingroup
\def\thefootnote{\fnsymbol{footnote}}\footnote[#1]{#2}\endgroup}

\begin{document}

\begin{titlepage}
\hfill\parbox[t]{4cm}{\texttt{FUT-Ph-03/15 \\
HU-EP-15/15 \\ 
ZMP-HH/15-5}}
\vspace{20mm}

\begin{center}

{\Large \bf Orbit method quantization of the AdS$_2$ superparticle}

\vspace{30pt}

\normalsize
{Martin Heinze,$^a$ Ben Hoare,$^b$ George Jorjadze,$^{b,\,c}~$ Luka Megrelidze$^d$
}
\\[6mm]

{\small
{\it\ ${}^a$II. Institut f{\"u}r Theoretische Physik, Universit{\"a}t Hamburg,\\
	Luruper Chaussee 149, 22671 Hamburg, Germany}\\[2mm]
{\it\ ${}^b$Institut f\"ur Physik und IRIS Adlershof, Humboldt-Universit\"at zu Berlin,\\
	Zum Gro\ss en Windkanal 6, 12489 Berlin, Germany}\\[2mm]
{\it${}^c$Razmadze Mathematical Institute and FU of Tbilisi,\\
	University Campus of Digomi, \\
	Agmashenebeli Alley 240, 0159, Tbilisi, Georgia}\\[2mm]
{\it${}^d$Ilia State University 
K. Cholokashvili Ave 3/5, 0162, Tbilisi, Georgia}\\[5mm]
\texttt{martin.heinze@desy.de}\vphantom{\{}\\[.7mm]
\{\texttt{ben.hoare, george.jorjadze}\}\texttt{@physik.hu-berlin.de}\\
\texttt{luka.megrelidze.1@iliauni.edu.ge}
}

\vspace{35pt}

\end{center}

\centerline{{\bf{Abstract}}}
\vspace*{5mm}
\noindent
\small
We consider the Hamiltonian reduction and canonical quantization of a massive AdS$_2$ superparticle realized on the coset  OSP$(1|2)/$SO$(1,1)$.
The phase space of the massive superparticle is represented as a coadjoint
orbit of a timelike element of $\mathfrak{osp}(1|2)$. This orbit has a well defined
symplectic structure and the OSP$(1|2)$ symmetry is realized as the Poisson
bracket algebra of the Noether charges.
We then construct canonical coordinates given by one bosonic and one fermionic oscillator, whose quantization leads to the Holstein-Primakoff type realization of $\mathfrak{osp}(1|2)$.
We also perform a similar analysis and discuss new features and inconsistencies in the massless case.

\normalsize
\vspace{15pt}

\end{titlepage}

\newpage

{\hypersetup{linkcolor=black}
\tableofcontents}

\section{Introduction}
	The quantization of constrained systems is an important problem of modern physics.
	Certainly a good motivation to be interested in this question is the $\AdS$/CFT correspondence\cite{Maldacena:1997re, Witten:1998qj, Gubser:1998bc}, which connects superstring theory in $d$ dimensional Anti-de-Sitter space ($\AdS_{d}$) to a super conformal field theory (CFT) on the $d-1$ dimensional conformal boundary.
	For the best studied example, the duality between the type IIB superstring on $\AdSfiveS$\cite{Metsaev:1998it} and $\calN = 4$ super Yang-Mills theory, there has been significant progress during the last decade \cite{Arutyunov:2009ga, Beisert:2010jr}, which can be attributed to existence of integrability in the planar limit \cite{Minahan:2002ve, Bena:2003wd}.
	In particular, due to the conjectured quantum integrability, powerful methods have been devised \cite{Arutyunov:2009zu, Arutyunov:2009ur, Bombardelli:2009xz, Gromov:2009bc, Gromov:2009tv, Gromov:2013pga}, which in principle allow one to predict the spectrum $E$
	of arbitrary string states at large 't Hooft coupling, $\l = (2\pi R^2 T_0)^2 \gg 1$, see {\it e.g.} \cite{Gromov:2009zb, Frolov:2010wt, Gromov:2011bz, Frolov:2012zv}. This ostensibly amounts to quantization of the system.

	With all these advances it is worth noting that the quantization of $\AdS$ superstrings from first principles is still an open question. For type IIB supergravity on $\AdSfiveS$, corresponding to the subsector of half BPS string states, the spectrum has been known for a long time \cite{Kim:1985ez, Gunaydin:1984fk}.
	In \cite{Horigane:2009qb} it was observed that the reformulation and first quantization of $\AdSfiveS$ supergravity in terms of a light-cone scalar superfield \cite{Metsaev:1999gz} is equivalent to quantizing the massless $\AdSfiveS$ superparticle. Hence, building on the phase space formalism of \cite{Metsaev:2000yu}, the resulting quantization indeed demonstrated a matching of the spectra, see also the related work \cite{Siegel:2010gm}. Indeed,
	following  the standard literature it seems that a thorough understanding of the massless superparticle would be a useful prerequisite for tackling the superstring.

	For massive string states, since the pioneering works \cite{Berenstein:2002jq, Gubser:2002tv, Frolov:2002av, Frolov:2003qc}, the majority of study has concentrated on semiclassical string dynamics at large 't Hooft coupling, $\l \gg 1$. For these one relies on some of the $\psu(2,2|4)$ charges diverging as $\sqrt{\l}$, rendering the corresponding string state {\it long}, or {\it heavy}, $E \propto \sqrt{\l}\,$.
	In the BMN limit \cite{Berenstein:2002jq} the total angular momentum on $\Sph^5$ $J$ diverges. Corrections in $\frac{1}{2 P_+} = (E + J)^{-1}$ were computed in \cite{Callan:2003xr, Callan:2004uv, Callan:2004ev} and \cite{Frolov:2006cc}, in which the so-called uniform light-cone gauge \cite{Arutyunov:2004yx, Arutyunov:2005hd} proved to be convenient. This then led to the perturbative calculation of the scattering $S$-matrix in this limit \cite{Arutyunov:2006yd, Klose:2006zd, Klose:2007rz}.
	
	However, in the case of finite $\psu(2,2|4)$ charges, for which semiclassical string solutions become {\it short}, there have been considerable challenges calculating the string spectrum beyond the leading order \cite{Gubser:2002tv}, $E \propto {\l}^{1/4}$.
	The reason for this is that in this regime the perturbative expansion of the Lagrangian formally breaks down,
	which can be traced back to the particular scaling behavior of the string zero-modes \cite{Passerini:2010xc}.
	For the uniform light cone gauge \cite{Arutyunov:2004yx, Arutyunov:2005hd} this is related to $P_- = E - J \propto \l^{1/4}$ becoming infinite.
	This raises the question of whether there is a more useful gauge choice to treat these excitations, or if the computations can be done in a gauge invariant way.

	Using static gauge \cite{Jorjadze:2012iy} and working in bosonic $\AdS_5 \times \Sph^5$, a generalization of the pulsating string \cite{deVega:1994yz, Minahan:2002rc} was constructed in \cite{Frolov:2013lva}, which allowed for unconstrained string zero-modes. This so-called single-mode string showed classical integrability and invariance under the isometries ${\rm SO}(2,4)\times {\rm SO}(6)$.
	For the lowest string excitation, dual to a member of the Konishi supermultiplet, the first non-trivial quantum correction to the spectrum was indeed reproduced. The fact that \cite{Frolov:2013lva} benefited from works on the massive bosonic $\AdSxS$ particle \cite{Dorn:2005ja, Dorn:2010wt} suggests that for the superstring it would be worthwhile understanding not only the massless, but also the massive $\AdSxS$ superparticle.

	The single-mode string solution of \cite{Frolov:2013lva} is the ${\rm SO}(2,4)\times {\rm SO}(6)$ orbit of the pulsating string \cite{deVega:1994yz, Minahan:2002rc}, motivating the investigation of symmetry group orbits of other semiclassical solutions. This is also appealing as the Kirillov-Kostant-Souriau method of coadjoint orbits, see also the seminal works \cite{Witten:1987ty, Alekseev:1988ce}, leads to a quantization in terms of the symmetry generators, which is manifestly gauge-independent.
	This idea was explored in \cite{Heinze:2014cga}, where, concentrating on the bosonic case of $\AdSthreeS$, orbits of the particle and spinning string \cite{Frolov:2003qc} were investigated. In a very natural and succinct way, the quantization procedure gave rise to a Holstein-Primakoff realization for the isometry algebra \cite{Holstein:1940zp, Dzhordzhadze:1994np, Jorjadze:2012jk}, in agreement with previous results for the particle  \cite{Dorn:2005ja, Dorn:2010wt}, as well as consistent short and long string limits for the spinning string.

	The goal of this work is to generalize the coadjoint orbit method to the case of supergroups.
As a first step in this direction, motivated by the AdS/CFT correspondence, we will investigate the $\AdS_2$ superparticle.
	Recalling that $\AdS_2$ can be realized as the coset $\SU(1,1)/\SO(1,1)$, here we will focus on the simplest generalization of this coset, 
$\OSP(1|2)/\SO(1,1)$.  The superalgebra $\osp(1|2)$ \cite{Pais:1975hu}
is the basic, hence classical and simple, Lie superalgebra of lowest dimension  \cite{Frappat:1996pb}, and has bosonic subalgebra $\mathfrak{su}(1,1)$ as required.

	Specifying the coset does not determine the action uniquely. Motivated by applications to Green-Schwarz string models, we choose the one possessing $\k$-symmetry in the massless case. However, in this case the $\kappa$-symmetry transformations leave us with an insufficient amount of fermions and consequently the quantization will only be consistent for the massive case.

	In fact, non-critical type IIA superstring theory on $\OSP(1|2)/\SO(2)$ has been studied in \cite{Verlinde:2004gt}, where due to $\kappa$-symmetry and reparametrization invariance, all fermionic and bosonic fields decouple, leaving only a supersymmetric Calogero-Moser model. The related work \cite{Adam:2007ws} discussed strings on the coset $\OSP(2|2)/\SO(1,1)\times\SO(2)$, 
	while in \cite{Bandos:1999pq} the coset model on $\OSP(1|4)/\SO(1,3)$ has been investigated. 
Furthermore, there have also been works \cite{Hikida:2007sz, Giribet:2009eb} on WZNW models and topological strings for the coset $\OSP(1|2)/\SO(2)$. Apart from these, the present work should also be relevant in the context of gauge/string duality for $\AdS_2 \times \Sph^2$ \cite{Zhou:1999sm, Berkovits:1999zq}, see also \cite{Bellucci:2002va, Ivanov:2002tb} and the recent works \cite{Sorokin:2011rr, Cagnazzo:2011at, Cagnazzo:2012uq, Hoare:2014kma}.

	The paper is organized as follows.
	After setting up notation in section \ref{setup}, in section \ref{bosAdS2} we revise how the coadjoint orbit method works for the massive and massless bosonic $\AdS_2$ particle \cite{Alekseev:1988ce}.
	We then proceed to the $\AdS_2$ superparticle in section  \ref{AdS2super}, in which we discuss the action of the $\OSP(2|1)/\SO(1,1)$ coset model and its $\kappa$-symmetry transformations in the massless case. With this in mind, the coadjoint orbit method is then generalized to the massive and massless $\AdS_2$ superparticle.
	We conclude and give an outlook in section \ref{conclusion}.

\section{Notation and Conventions}\label{setup}

Using the Pauli matrices $\boldsymbol{\sigma}_j$ $(j=1,2,3)$, a basis of $\mathfrak{su}(1,1)$ can be written as
\begin{equation}\label{su(1,1) basis}
{\bf t}_0=-i{ \boldsymbol\s}_3~, \quad {\bf t}_1={\boldsymbol\s}_1~, \quad
{\bf t}_2={\boldsymbol\s}_2~.
\end{equation}
The matrices ${\bf t}_a$ $(a=0,1,2)$ satisfy the relations
\begin{equation}\label{tt=}
{\bf t}_a\,{\bf t}_b=\eta_{ab}\,{\bf I}+\epsilon_{ab}\,^{c}\,
{\bf t}_c~,
\end{equation}
where ${\bf I}$ is the unit matrix, $\eta_{ab}=\mbox{diag}(-1,1,1)$ and
$\epsilon_{abc}$ is the Levi-Civita tensor, with
$\epsilon_{012}=1$. The Killing  form defined by the normalized trace
$\langle\, {\bf t}_a\,{\bf t}_b\,\rangle\equiv\frac{1}{2}\,
\mbox{Tr}({\bf t}_a\,{\bf t}_b)
=\eta_{ab}$
provides the isometry between $\mathfrak{su}(1,1)$ and 3d Minkowski space,
since for a pair of $\mathfrak{su}(1,1)$ vectors ${\bf u}=u^a\,{\bf t}_a$
and ${\bf v}=v^a{\bf t}_a$, one has $\langle\, {\bf u}\,{\bf v}\,\rangle=u^a v_a$.
Expanding ${\bf v} \in \mathfrak{su}(1,1)$ as
\be\label{expansion of v}
{\bf v}=v_1\,{\bf t}_1 +v_+{\bf t}_- +v_-{\bf t}_+~,
\ee
with ${\bf t}_\pm=\frac{1}{2}({\bf t}_2\pm{\bf t}_0)$, one finds
$v_1=\langle\, {\bf v}\,{\bf t}_1\,\rangle$, $v_\pm=2\langle\,{\bf v}\,{\bf t}_\pm\,\rangle$ and
\be\label{v^2}
 \langle\, {\bf v}^2\,\rangle=v_1^2+v_+v_-~.
\ee

We consider the following matrix representation of the real superalgebra $\mathfrak{osp}(1|2)$
\begin{equation}\label{su'(1,1|1) basis}
T_a=\left( \begin{array}{cc}
 {\bf t}_a & 0 \\ 0 & 0
 \end{array}\right)~, ~~~~~~
 S_-=\left( \begin{array}{ccc}
 0 & 0 & 1\\ 0 & 0 & 1\\ -1 & 1 & 0\end{array}\right)~,
 ~~~~~~ S_+=-i\left( \begin{array}{ccc}
 0 & 0 & 1\\ 0 & 0 & -1\\ 1 & 1 & 0
 \end{array}\right)~,
\end{equation}
where ${\bf t}_a$ are given by \eqref{su(1,1) basis}.
The similarity transformation $U^{-1}\,T_a\, U$,  $U^{-1}\,S_\pm\, U$, with
\be\label{U}
U=\frac{1}{\sqrt 2}\left( \begin{array}{ccc}
 1 & i & 0\\ i & 1 & 0\\ 0 & 0 & \sqrt 2\end{array}\right)
\ee
maps \eqref{su'(1,1|1) basis} to the basis vectors of $\mathfrak{osp}(1|2)$
in the usual defining representation.

The commutation relations of the basis elements
\be\label{basis 1}
	T_1~, \qquad T_\pm =\frac{1}{2}(T_2 \pm T_0)~,  \qquad S_\pm
\ee
take the following  compact form
\be\ba\label{comm T1}
&[T_1,\,T_\pm]=\mp \,2T_\pm~,&    &[T_1,\,S_\pm]=\mp\, S_\pm~,&   &[T_\pm,\,S_\mp]=S_\pm~,& ~[T_\pm,\,S_\pm]=0 ~,\\
&[T_-,\,T_+]=T_1~,&    &[S_-,\,S_+]_+=-2iT_1~,&       &[S_\pm,\,S_\pm]_+=\mp\,4i\,T_\pm~.&
\ea\ee

	The normalized supertrace  $\langle {\mathfrak a}\, \mathfrak b \rangle=
\frac{1}{2}\big(({\mathfrak a}\, \mathfrak b)_{11}+({\mathfrak a}\, \mathfrak b)_{22}-
({\mathfrak a}\, {\mathfrak b})_{33}\big)$ defines a Killing form in
	$\mathfrak{osp}(1|2)$ with nonzero components
\be\label{supertaces}\B{
\langle T_1\,T_1 \rangle=1~, \quad \langle T_+\,T_- \rangle=\langle T_-\,T_+ \rangle=\frac{1}{2}~,
\quad \langle S_+\,S_- \rangle=- \langle S_-\,S_+ \rangle=2i~.}
\ee
	Then expanding $V\in \mathfrak{osp}(1|2)$ in the basis \eqref{basis 1}
\be\label{expansion of V}
V=V_1 T_1+V_+\,T_- + V_-\,T_+ +V^s_+ \,S_- +V^s_- \,S_+~,
\ee
	one finds $V_1=\langle V\,T_1 \rangle$, $V_\pm=2\langle V\,T_\pm \rangle$ and $V^s_\pm=\pm\frac{i}{2}\langle V\,S_\pm \rangle.$

\section{Coset construction of the bosonic \texorpdfstring{AdS$_2$}{AdS2} particle}
\label{bosAdS2}

\subsection{Classical description}
	Let us start by considering particle dynamics on AdS$_2$, as described by
the coset sigma model for $\mbox{SU}(1,1)/\mbox{SO}(1,1)$. Explicitly we use
the basis for $\mbox{SU}(1,1)$ given in section \ref{setup} and consider the
$\mbox{SO}(1,1)$ gauge transformation $g(\tau)\mapsto  e^{\a(\tau){\bf t}_1}\,
g(\tau)$ generated by ${\bf t}_1$. The corresponding gauge invariant action is
\be\label{gauged SL(2,R) action}
S=\int \mbox{d}\tau\,\left[\frac{\langle\, (\dot{g}\,g^{_{-1}}-A{\bf t}_1)^2\,\rangle}{2\xi}
-2\xi \m^2\right]~,
\ee
where $\xi$ is the worldline einbein and $A$ transforms as a gauge potential $A(\tau)\mapsto A(\tau)+\dot\a(\tau).$ Varying \eqref{gauged SL(2,R) action} with respect to $A$
gives $A=\langle\,{\bf t}_1\, \dot{g}\, g^{_{-1}}\, \rangle$ and its insertion back in \eqref{gauged SL(2,R) action}
leads to the gauge invariant action written solely in terms of $g$
	\be\label{gauged SL(2,R) action 1}
		S=\int \mbox{d}\tau\,\left[\frac{\langle\,(\dot{g}\,g^{_{-1}})^2 \,\rangle-
		\langle\, \dot{g}\, g^{_{-1}}\,{\bf t}_1\, \rangle^2}{2\xi} -2\xi \m^2\right]~.
	\ee
	Defining ${\bf v} = \dot g g^{_{-1}}$ and using \eqref{expansion of v} and \eqref{v^2},
	the action \eqref{gauged SL(2,R) action 1} can be written as
	\be\label{gauged SL(2,R) action 1'}
	S=\int \mbox{d}\tau\,\left[\frac{v_+\,v_-}{2\xi} - 2\xi \m^2\right]~.
	\ee
	The gauge transformation of ${\bf v}$ is given by ${\bf v}\mapsto e^{\a{\bf t}_1}\, {\bf v}\,e^{-\a{\bf t}_1}+\dot\a\,{\bf t}_1$.
	Using $[{\bf t}_1, {\bf t}_\pm]=\mp \,2{\bf t}_\pm$, we obtain
${v}_\pm\mapsto e^{\pm 2\a}\,{v}_\pm$, which explicitly demonstrates
the gauge invariance of \eqref{gauged SL(2,R) action 1'}.

The action \eqref{gauged SL(2,R) action 1'} can also be written in terms of a Lie algebra valued gauge invariant variable ${\bf x}=g^{_{-1}}\,{\bf t}_1 g $.
Indeed, one has $\dot{\bf x}= g^{_{-1}}\,[{\bf t}_1, {\bf v}]g$ and $[{\bf t}_1, {\bf v}]=2{v}_+{\bf t}_- -2{v}_-{\bf t}_+$.
Hence, $\langle\,\dot {\bf x}^2\,\rangle=-4{v}_+{v}_-$ and  \eqref{gauged SL(2,R) action 1'} becomes
	\be\label{AdS_2 action}
	S=\int \mbox{d}\tau\,\left[-\frac{\dot x_a\dot x^a}{2\tilde\xi} - \frac{\tilde\xi\, \mu^2}2 \right]~.
	\ee
Here, $\tilde \xi=4\xi$ and $x_a=\langle\,{\bf  x}\,{\bf t}_a \,\rangle$ are the coordinates of ${\bf x}$ in the basis \eqref{su(1,1) basis}.
These coordinates are real and they are bounded on the hyperboloid $-x_a\,x^a=(x_0)^2-(x_1)^2-(x_2)^2=-1$, since $\langle\,{\bf  x}^2\rangle=1$.
	The time coordinate corresponds to the polar angle in the $(x_1,x_2)$ plane, and hence,
after considering the universal cover, the action \eqref{AdS_2 action} describes the AdS$_2$ particle
with mass $\mu$.
	This can be seen explicitly by introducing the global coordinates
\begin{equation}
x_1 + i x_2 = \cosh \rho \, e^{-i t} ~ , \qquad x_0 = \sinh \rho ~ , \qquad g = \exp\big(\frac{\l\,{\bf t}_1}2\big) \exp\big(\frac{\rho\,{\bf t}_2}2\big) \exp\big(\frac{t \,{\bf t}_0}{2}\big)~,
\end{equation}
such that action \eqref{AdS_2 action} becomes
\be\label{AdS_2 actioncoords}
S=\int \mbox{d}\tau\,\left[\frac{-\cosh^2 \rho \, \dot t^2 + \dot \rho^2 }{2\tilde\xi} -\frac{\tilde\xi\, \m^2}{2}\right]~.
\ee

The global symmetry of \eqref{gauged SL(2,R) action 1} is given by the right multiplications $g\mapsto g\,h$,
with $h\in \mbox{SU}(1,1)$,
and the corresponding Noether charge reads
\be\label{Charge}
R=\frac{g^{_{-1}}\,\dot{g}-\langle\,\dot{g}\, g^{_{-1}}\, {\bf t}_1\, \rangle\, g^{_{-1}}\,{\bf t}_1\, g}{\xi}~.
\ee
Writing $R$ in the form
\be\label{R}
R=2\left( \begin{array}{cr}
 -iE & -iB\\iB^*& iE
 \end{array}\right)~,
\ee
with $B=B_2+iB_1$ and $B^*=B_2-iB_1$, we find that $E$ corresponds to the particle energy, while $B_1$ and $B_2$
are the boost generators.

	Varying \eqref{gauged SL(2,R) action 1} with respect to $\xi$ gives the mass-shell condition $\langle\,R\,R\,\rangle + 4 \m^2=0$, which is equivalent to the Casimir number relation
	\be\label{Casimir 1}
	E^2-B^*\,B=\mu^2~.
	\ee

	It is interesting to note that the massive model considered here is classically equivalent to a massless particle moving on AdS$_2 \times$ S$^1$ with fixed angular momentum on S$^1$. To see this, we extend the $\mu = 0$ case of \eqref{gauged SL(2,R) action} as follows
	\be\label{gauged SL(2,R) action plus u1}
		S=\int \mbox{d}\tau\,\left[\frac{\langle\, (\dot{g}\,g^{_{-1}}-A{\bf t}_1)^2\,\rangle + \dot \phi^2}{2\xi}\right]~,
	\ee
	where $\phi$ is the angle on S$^1$.
	Varying with with respect to $\xi$ gives the mass-shell condition $\langle\, (\dot{g}\,g^{_{-1}}-A{\bf t}_1)^2\,\rangle + \dot \phi^2 = 0\,$.
Furthermore, cyclicity of $\phi$ yields the integral of motion $\frac{\dot\phi}{\xi} = 2\mu\,$, which when inserted into the mass-shell condition leads us back to \eqref{Casimir 1}.

\subsection{First order formulation 
and 
quantization}

Applying the Faddeev-Jackiw  formalism to \eqref{gauged SL(2,R) action 1}, one finds the first order action
\be\label{First order action}
S=\int \mbox{d}\tau\,\left[\langle L\,\dot{g}\,g^{_{-1}}\,\rangle
- \frac{\xi}{2}
\left(L_+L_-+4\m^2 \right)-A L_1\right]~,
\ee
where $L$ is a Lie algebra valued phase space variable and $L_1$, $L_\pm$ are its components as in \eqref{expansion of v}.
The variables $\xi$ and $A$ now play the role of Lagrange multipliers and their variations provide the constraints
\be\label{constraints}
L_+L_-+4\m^2=0~, \qquad\qquad  L_1=0~.
\ee

Thus, the system is described by the 1-form and the Noether charge
\be\label{Theta and R}
\Theta=\langle\,L\,\mbox{d}g\,g^{_{-1}}\,\rangle~, \qquad \qquad R=g^{_{-1}}\,L\,g~,
\ee
restricted to the constraint surface \eqref{constraints}.
	The reduction schemes for the massive and the massless cases are different and hence we analyze them separately.

	First we consider the massive case, for which one can use the parametrization $L_\pm = \mp 2\m\,e^{\pm\g}$,
	for some $\g$, and hence $L$ can be written as $L=2\m\,e^{\frac{\g}{2}\,{\bf t}_1}\, {\bf t}_0\,e^{-\frac{\g}{2}\,{\bf t}_1}$.
	Setting $g=e^{\frac{\g}{2}\,{\bf t}_1}\,g_r$,
	\eqref{Theta and R} then takes a coadjoint orbit form \cite{Alekseev:1988ce}
\be\label{Theta and R 1}
\Theta=2\m\langle\,{\bf t}_0\,\mbox{d}g_r\,g^{_{-1}}_r\,\rangle~, \qquad \qquad R=2\m\,g^{_{-1}}_r\,{\bf t}_0\,g_r~.
\ee
	With the parametrization
\be\label{SU(1,1) parametrization}
g_r={e^{\phi\,{\bf t}_0}}\left( \begin{array}{cr}
 \sqrt{1+z^* z} & z~~~~~\\z^*& \sqrt{1+z^* z}
 \end{array}\right)~,
\ee
equation \eqref{Theta and R 1} reduces to
\be\label{Theta and R 2}
\Theta=
\frac{i}{2}\,(b^*\mbox{d}b-b\mbox{d}b^*)-2\m\mbox{d}\phi~, \qquad
	R=2\left( \begin{array}{cr}
 -i(\m+b^* b) & -i\sqrt{2\m+b^* b}\,\, b  \\i b^*\sqrt{2\m+b^* b}& i(\m+b^* b)
 \end{array}\right),
\ee
where $b=\sqrt{2\m}\,\,z$ and $b^*=\sqrt{2\m}\,\,z^*$. From \eqref{R} we can then extract the Noether charges
\be\label{DI=}
E=\mu+b^* b~,
\qquad B= \sqrt{2\m+b^*b}\,\,b~,
\qquad B^*=b^*\sqrt{2\mu+b^*b}~.
\ee

The symplectic form $\Omega=\mbox{d}\Theta$ obtained from \eqref{Theta and R 2} is canonical
$\Omega=i\,\mbox{d}b^*\wedge \mbox{d}b$ and in terms of $(B, B^*)$ it takes the Kirillov-Kostant  form
\be\label{2 Form}
\Omega=i\frac{\mbox{d}B^* \wedge \mbox{d}B}{2\sqrt{\m^2+B^*B}}=\frac{\mbox{d}B_1 \wedge \mbox{d}B_2}{E}~.
\ee
Quantization in the oscillator variables $(b, b^*)$ then leads to the Holstein-Primakoff
realization of the symmetry generators \eqref{DI=} with Casimir number
\be\label{Casimir}
C=E^2-\frac{1}{2}( B^*B+B B^*)=\m(\m-1).
\ee
This representation is unitary and irreducible for $\m>0$.

Now we consider the massless case. Note that at $\m=0$ the 2-form \eqref{2 Form} is singular at the origin $B_1=0=B_2$.
This point corresponds to the massless particle with
zero energy and should be removed from the phase space as for Minkowski space. From \eqref{constraints} we then have two possibilities,
either $L_+=-e^{\g_+}$ and $L_-=0$, or $L_+=0$ and $L_-=e^{-\g_-}$. Let us analyze the second one,
which corresponds to $L=e^{\frac{\g}{2}\,{\bf t}_1}\, {\bf t}_+\,e^{-\frac{\g}{2}\,{\bf t}_1}$.
As in the massive case, setting $g=e^{\frac{\g}{2}\,{\bf t}_1}\,g_r$ yields
\be\label{massless data}
\Theta=\langle {\bf t}_+\,\mbox{d}{g}_r\,g^{_{-1}}_r\rangle~, \qquad R=g^{_{-1}}_r\,{\bf t}_+\,g_r~,
\ee
and using the parametrization $g_r=e^{2\a {\bf t}_+}\,e^{\frac{\b}{2}\,{\bf t}_1}\,e^{2\g {\bf t}_-}$, we obtain
\be\label{massless data 1}
\Theta=e^\b\mbox{d}\g~,
\qquad R=e^\b({\bf t}_+ -2\g {\bf t}_1-4\g^2{\bf t}_-)~.
\ee
The dynamical integrals are then given by
\be\label{massless DI}
E=(\g^2+1/4)e^\b~, \qquad B_1=-\g\,e^\b~, \qquad B_2=-(\g^2-1/4)e^\b~,
\ee
and $\Omega=\mbox{d}\Th=\mbox{d}e^\b\wedge\mbox{d}\g$ again takes the Kirillov-Kostant form \eqref{2 Form} for $\m=0$.
The case $L_+=-e^{\g_+}$ and $L_-=0$ gives the same answer in a similar way.

The dynamical integrals \eqref{massless DI} can be expressed through canonical oscillator variables,
\be\label{massless DI=}
E=b^* b~, \qquad B= \sqrt{b^*b}\,\,b~, \qquad B^*=b^*\sqrt{b^*b}~.
\ee
with $|b|>0$, and one arrives again at the Holstein-Primakoff representation for $\mu=0$.
This representation becomes irreducible
if one removes the ground state $|\,0\,\rangle$, which is annihilated by all symmetry generators.
Note that the resulting representation is unitary equivalent to the representation \eqref{DI=} for $\mu=1$.

\section{Coset construction of the
\texorpdfstring{AdS$_2$}{AdS2} superparticle} \label{AdS2super}

\subsection{Classical description}
	Let us consider $g(\tau)\in \mbox{OSP}(1|2)$ and the gauge transformations $g(\tau)\mapsto e^{\a(\tau) T_1}\,g(\tau)$.
	The 'left current' $V=\dot g\,g^{_{-1}}$ then transforms as $V\mapsto e^{\a T_1}\,V\,e^{-\a T_1}+\dot\a\,T_1$ and, 
	using the expansion \eqref{expansion of V}, we find the following gauge transformations for its components
	\be\label{tr of jpm}
		V_1\mapsto V_1+\dot\a~, \qquad V_\pm \mapsto e^{\pm2\a}\,V_\pm \qquad V^s_\pm \mapsto e^{\pm \a}\,V^s_\pm~.
	\ee

We describe a superparticle on AdS$_2$ by the following gauge invariant action
\be\label{gauged SL(2,R) action s-p}
S=\int \mbox{d}\tau\,\left[\frac{V_+\,V_-}{2\xi}-2\xi\,\m^2\right]~.
\ee
The Noether charge related to the right multiplications $g\mapsto \,g\, h$ is then given by
\be\label{R s-p}
R=\frac{g^{_{-1}}(V_+\,T_- +V_-\,T_+)g}{\xi}~,
\ee
and it satisfies the mass-shell condition $\langle R\,R\rangle+4\m^2=0$. 

	The form of this action is motivated by the supercoset formulation of the
Green-Schwarz string on AdS$_5 \times$ S$^5$ \cite{Metsaev:1998it}, AdS$_2
\times$ S$^2$ \cite{Berkovits:1999zq}, $\AdS_2$ with NS-NS \cite{Verlinde:2004gt} or R-R flux \cite{Adam:2007ws}, and other integrable AdS backgrounds \cite{Zarembo:2010sg, Wulff:2014kja}.
	Removing the dependence of the spacelike
	worldsheet coordinate the WZ term of those actions drops out and in all cases we are just left with the square of the current projected onto the Grassmann even part of the coset.
	Furthermore, in the massless case this implies that this action will have a
$\kappa$-symmetry halving the number of fermionic degrees of freedom. Indeed, in this case the action
coincides with a truncation of the AdS$_2 \times$ S$^2$ super 0-brane action
constructed in \cite{Zhou:1999sm}.

It is also of interest to look at the explicit form of this action. This
is readily doable as there are only two Grassmann odd generators in
$\mathfrak{osp}(1|2)$ and hence only two fermionic fields. Parametrizing
the gauge fixed group field as
\begin{equation}
g= \exp\big(\frac{\psi_1 S_+}2 + \frac{\psi_2 S_-}{2}\big) \exp\big(\frac{\rho T_2}2\big) \exp\big(\frac{t T_0}2\big) ~,
\end{equation}
we find the following action
\begin{equation}\label{act1}
S = \int \mbox{d}\tau\,\left[\frac{(1+ i\psi_1\psi_2)(\cosh\rho\,\dot t + \dot \rho-i\psi_1\dot\psi_1)(-\cosh\rho\,\dot t+\dot \rho+i\psi_2\dot\psi_2)}{2\tilde \xi}
-\frac{\tilde \xi\,\mu^2}2\right]~,
\end{equation}
where $\tilde \xi = 4\xi$.

To take the flat space limit we set
\begin{equation}
t \to \frac {x^0}R~, \qquad \rho \to \frac {x^1} R~, \qquad \psi_{1,2} \to \frac{\chi_{1,2}}{\sqrt{R}}~, \qquad \tilde \xi \to \frac{\tilde \xi}{R^2} ~ , \qquad \mu \to \mu R ~,
\end{equation}
and take $R \to \infty$. Doing so we find
\begin{equation}\label{act2}
S = \int \mbox{d}\tau\,\left[\frac{(\dot x^0 + \dot x^1-i\chi_1\dot\chi_1)(-\dot x^0+\dot x^1+i\chi_2\dot\chi_2)}{2\tilde \xi}
-\frac{\tilde \xi\,\mu^2}2\right]~,
\end{equation}
which can easily be seen to be equivalent to the $2d$, $\mathcal N =(1,1)$
superparticle \cite{Brink:1981nb}
\begin{equation}\label{flat}
S = \int \mbox{d}\tau\,\left[\frac{\eta_{\mu\nu}(\dot x^\mu - i \bar \chi \gamma^\mu \dot\chi)(\dot x^\nu-i\bar \chi\gamma^\nu\dot \chi)}{2\tilde\xi} - \frac{\tilde\xi\,\mu^2}2\right]~.
\end{equation}
the field content of which is given by two bosons and one Majorana spinor
\begin{equation}
x =
\begin{pmatrix} x^0 \\ x^1\end{pmatrix}~, \qquad
\chi =\frac{1}{\sqrt{2}} \begin{pmatrix}\chi_1\\\chi_2\end{pmatrix}~, \qquad \eta = \text{diag}(-1,1)~,\quad \gamma^0 = \sigma_2~,\quad \gamma^1 = i \sigma_1~.
\end{equation}

Considering fluctuations around a non-trivial bosonic background,
the actions \eqref{act1} and \eqref{act2} describe two bosonic
fields satisfying second order equations of motion, and two fermionic fields
satisfying first order equations of motion. Therefore, in the massive case,
taking account of the mass-shell condition, we find one on-shell bosonic and
one on-shell fermionic degree of freedom.

In the massless case, due to the presence $\kappa$-symmetry halving the number
of fermionic degrees of freedom, we encounter a problem. The
$\kappa$-symmetry can be most easily seen if we consider the following
field redefinitions in \eqref{act1} with $\mu=0$
\begin{equation}\begin{split}
\bar \xi  = \operatorname{sech}^2\rho (1-i\psi_1\psi_2)\tilde\xi \ , \qquad
x^+ & = t + 2\tan^{-1} e^\rho \ , \qquad x^- = t - 2 \tan^{-1}e^{\rho} \ ,
\\ \chi_1 & = \sqrt{\operatorname{sech}\rho}\,\psi_1\ , \hspace{36pt}
\chi_2 = \sqrt{\operatorname{sech}\rho}\,\psi_2\ .
\end{split}\end{equation}
The resulting action is given by
\begin{equation}\label{act3}
S= \int \mbox{d}\tau\,\left[-\frac{(\dot x^+ - i \chi_1\dot\chi_1)(\dot x^- - i \chi_2\dot\chi_2)}{2\bar \xi}\right]\ ,
\end{equation}
which remarkably is formally equivalent to the massless case of the $2d$, $\mathcal{N}=(1,1)$
action \eqref{act2} when we take $x^\pm = x^0 \pm x^1$ 
and $\bar \xi = \tilde \xi$.
The action \eqref{act3} is then invariant under the following $\kappa$-symmetry transformation \cite{Siegel:1983hh}
\begin{equation}\begin{split}\label{kappasym}
\delta \hat{\xi} = -2 i \bar{\xi} (\dot\chi_1 \kappa_2 + \dot\chi_2 \kappa_1) \ , \qquad
\delta x^+ & = - i \chi_1 P^+ \kappa_2 \ , \qquad
\delta x^- = - i \chi_2 P^- \kappa_1 \ ,
\\
\delta \chi_1 & = - P^+ \kappa_2 \ , \hspace{39.5pt}
\delta \chi_2 = -P^- \kappa_1 \ ,
\end{split}
\end{equation}
where $\kappa_{1,2}$ are infinitesimal Grassmann odd parameters that are allowed to depend on $\tau$
and we have defined
\begin{equation}
P^+ = \dot x^+ - i \chi_1\dot\chi_1 \ , \qquad P^- = \dot x^- - i \chi_2\dot \chi_2 \ .
\end{equation}
Note that the mass-shell condition following from \eqref{act3} is $P_+ P_- = 0$ and
hence the on-shell rank of the $\kappa$-symmetry transformations \eqref{kappasym} is one.

As we will see, this problem will reappear when
we try to quantize the massless AdS$_2$ superparticle based on the action \eqref{gauged SL(2,R) action s-p} and the supergroup
OSP$(1|2)$. Indeed, the fact
that we do not have enough fermionic degrees of freedom is a consequence of the
fact that we started with the superalgebra $\mathfrak{osp}(1|2)$, which has only
two fermionic generators. To properly treat the massless superparticle on
AdS$_2$ we should instead start with the superalgebra $\mathfrak{su}(1,1|1) \simeq \osp(2|2)$,
gauging an $\mathfrak{so}(1,1) \oplus \mathfrak{u}(1)$ subalgebra
\cite{Adam:2007ws}, which has twice the number of fermionic generators.

\subsection{First order formulation}

In the first order formalism  \eqref{gauged SL(2,R) action s-p} is equivalent to
\be\label{First order action s-p}
S=\int \mbox{d}\tau\,\left[\langle L\,\dot{g}\,g^{_{-1}}\,\rangle
- \frac{\xi}{2}\left(L_+L_-+4\m^2 \right)-A_1 L_1-A^s_+\,L^s_--A^s_-\,L^s_+\right]~,
\ee
where $L_1$, $\,L_\pm$, $\,L^s_\pm$ are the components of $L$ in the basis \eqref{basis 1},
($~\xi$, $A_1$, $A^s_\pm$) play the role of Lagrange multipliers and their variations  give the constraints
\be\label{constraints s-p}
L_+L_-+4\m^2=0~, \qquad  L_1=0~, \qquad L^s_\pm=0~.
\ee

As in the bosonic case, we have the 1-form $\Theta=\langle\,L\,\mbox{d}g\,g^{_{-1}}\,\rangle$ and the Noether charge $\,R=g^{_{-1}}\,L\,g\,$,
	only now $g\in \mbox{OSP}(1|2)$, $L\in \mathfrak{osp}(1|2)$ and the system has to be reduced to
the constraint surface \eqref{constraints s-p}. Similarly to the bosonic case, $L$ and $g$ can then be parameterized $L=2\m\,e^{\frac{\g}{2}\,T_1}\, T_0\,e^{-\frac{\g}{2}\,T_1}$,
$\,g=e^{\frac{\g}{2}\,T_1}\,g_r$, which leads to
\be\label{Theta and R S1}
\Theta=2\m\langle\,T_0\,\mbox{d}g_r\,g^{_{-1}}_r\,\rangle~, \qquad \qquad R=2\m\,g^{_{-1}}_r\,T_0\,g_r~.
\ee

To find a suitable parametrization of $g_r$ we represent it as $g_r=g_f\, g_b$, with $g_f$ purely fermionic and $g_b$ purely bosonic.
Setting $g_f=e^{\th_+S_- + \th_-S_+}$, where $S_\pm$ are the fermionic generators in \eqref{su'(1,1|1) basis} and $\th_\pm$ are real Grassmann odd parameters, we find
\be\label{gf}
g_f=\left( \begin{array}{ccc}
 1+\frac{\th^* \th}{2} & 0& \th\\0& 1+\frac{\th^* \th}{2}& \th^*\\
 -\th^* &  \th & 1-\th^* \th
 \end{array}\right),
\ee
with $\th=\th_+-i\th_-$, $\th^*=\th_++i\th_-$.
Similarly to \eqref{SU(1,1) parametrization}, $g_b$ is chosen as
\be\label{gb}
g_b=e^{\phi\,T_0}\left( \begin{array}{ccc}
 \sqrt{1+u^* u} & u& 0\\u^*& \sqrt{1+u^* u}& 0\\
 0 &  0 & 1
 \end{array}\right),
\ee
and the product of \eqref{gf} and \eqref{gb}  can be written as
\be\label{SU(1,1.1) parametrization}
g_r=e^{\phi\,T_0}\left( \begin{array}{ccc}
 \sqrt{1+z^* z}+\frac{\psi^*\psi}{2\sqrt{1+z^* z}} & z& \psi\\z^*& \sqrt{1+z^* z}+
 \frac{\psi^*\psi}{2\sqrt{1+z^* z}}& \psi^*\\
 z^*\psi-\sqrt{1+z^*z}\,\psi^* &  \sqrt{1+z^*z}\,\psi -z\psi^*  & 1-\psi^*\psi
 \end{array}\right)~,
\ee
where $\psi=\th\,e^{i\phi}$, $z=u\left(1+\frac{\th^* \th}{2}\right)$ and $\psi^*$, $z^*$ are their
complex conjugations.

The inverse to \eqref{SU(1,1.1) parametrization} is given by
\be\label{SU(1,1|1) parametrization1}
g_r^{_{-1}}=\left( \begin{array}{ccc}
 \sqrt{1+z^* z}+\frac{\psi^*\psi}{2\sqrt{1+z^* z}} & -z&z\psi^* - \sqrt{1+z^*z}\,\psi\\-z^*& \sqrt{1+z^* z}+\frac{\psi^*\psi}{2\sqrt{1+z^* z}}& z^*\psi -\sqrt{1+z^*z}\,\psi^*\\
 \psi^*&-\psi  & 1-\psi^*\psi
 \end{array}\right)e^{-\phi\,T_0}~,
\ee
and then the 1-form and the Noether charge in \eqref{Theta and R S1} become
\be\label{R=3}
\Th=\frac{i}{2}(b^*\mbox{d}b-b\,\mbox{d}b^*)+\frac{i}{2}(f^*\mbox{d}f+f\,\mbox{d}f^*)-2\m\,\mbox{d}\phi~,~~~
R=\left( \begin{array}{ccc}
 -2iE & -2i B& -i F\\2i B^*& 2iE& i F^*\\
 -i F^*&  -iF  & 0
 \end{array}\right),
 \ee
where $b=\sqrt{2\m}\,\,z,$ $\,b^*= \sqrt{2\m}\,\,z^*,$ $\,f=\sqrt{2\m}\,\, \psi$,
$\,f^*=\sqrt{2\m}\,\, \psi^*$ are canonical coordinates and the matrix elements in \eqref{R=3} read
\bea\nonumber
&&E=\m+b^* b+\frac{f^* f}{2}\\  \label{b,f to B,F}
&&B=\sqrt{2\m+b^* b}\,\,b+\frac{f^* f}{2\sqrt{2\m+b^* b}}\,\,b  \qquad B^*=(B)^*~,\\  \nonumber
&&F=\sqrt{2\m+b^* b}\,f+f^*\,b~, \qquad  \qquad ~~~~~ F^*=(F)^* ~.
\eea

The canonical coordinates define the Poisson brackets
\be\label{Can-PB}
\{A_1,\,A_2\}=i\left(\frac{\p A_1}{\p b^*}\frac{\p A_2}{\p b}-\frac{\p A_1}{\p b}\frac{\p A_2}{\p b^*}\right)
-i\left(\frac{\overleftarrow{\p} A_1}{\p f^*}\frac{\overrightarrow{\p} A_2}{\p f}+\frac{\overleftarrow{\p} A_1}{\p f}\frac{\overrightarrow{\p} A_2}{\p f^*}\right)~,
\ee
where $\overrightarrow{\p}$ and $\overleftarrow{\p}$ denote the left and the right derivatives, respectively.
The Poisson brackets of the dynamical integrals \eqref{b,f to B,F} satisfy the $\mathfrak{osp}(1|2)$ algebra
\be\ba\label{psu(1,1.1) algebra by PB}
	&\{E,B\}= i\, B~,&  \qquad  &\{E,B^*\}= -i\, B^*~,&  \qquad &\{B,B^*\}=-2i E~,&\\
	&\{E,F\}=\f{i}{2}\,F~,&             &\{B, F\}=0~,&                  &\{B^*, F \}=i F^* ~,& \\
	&\{E,F^*\}=-\f{i}{2}\,F^*~,&      &\{B, F^*\}=-i\,F~,&                   &\{B^*, F^* \}=0 ~,& \\
	&\{F , F^*\}=-2i E~,&          &\{F ,F\}=-2i\,B~,&       &\{F^* ,F^*\}=-2i\,B^*~.&
\ea\ee
The Casimir number obtained from \eqref{b,f to B,F} corresponds to the mass-square
\be\label{Casimir 2}
C=E^2- B^* B-\frac{1}{2}\,F^*F=\m^2~,
\ee
and the energy given in terms of other symmetry generators is
\be\label{E=}
E=E_B +\frac{F^* F}{4 E_B}~, \qquad \mbox{with} \quad E_B=\sqrt{\m^2+B^* B}~.
\ee
From \eqref{b,f to B,F} we also find
\be\label{F*F}
F^* F=2\m f^* f~,
\ee
which allows us to invert the map from  $(b, b^*, f, f^*)$ to $(B, B^*, F, F^*)$
\be\ba\label{B,F to b,f}
&b=\left(1-\frac{F^* F}{8\m E_B}\right)\frac{B}{\sqrt{\m+ E_B}}~,\qquad \qquad && b^*=(b)^*~ ,
\\
&f=\frac{\sqrt{\m+ E_B}\,\,F}{2\m}-\frac{BF^*}{2\m \sqrt{\m+ E_B}}~,   && f^*=(f)^*~.
\ea\ee
Using then the coordinates $\xi^k=(B, B^*, F, F^*)$, we can write the canonical 1-form
$\Th=\frac{i}{2}(b^* \mbox{d}b -b\,\mbox{d}b^*)+\frac{i}{2}(f^* \mbox{d}f +f\,\mbox{d}f^*)$ as follows
$$\Th=\Th_B\, \mbox{d}B+\Th_{B^*}\,\mbox{d}B^*+\Th_F\,\mbox{d}F+\Th_{F^*}\,\mbox{d}F^*,$$
with
\be\ba\label{1-form coefficients}
&\Th_B=\frac{i}{4\m^2}\left(\frac{2\m^2}{E_B+\m}-\frac{F^* F}{2E_B}\right)B^*~, \qquad  & \Th_{B^*}=\left(\Th_B\right)^*~,
\\
&\Th_F= \frac{i}{4\m^2}( E_B F^*-B^*F)~, \qquad  & \Th_{F^*}=-\left(\Th_F\right)^*.
\ea\ee
The matrix elements for the symplectic form $\Omega=\mbox{d}\Th$
are given by
\be\ba\label{symplectic matrix}
	&\Omega_{B B}=\Omega_{B^* B^*}=0~,  
		&&\Omega_{B B^*}=-\Omega_{B^* B}=\p_B\Th_{B*}-\p_{B^*}\Th_{B}~, \\
	&\Omega_{B F}=\Omega_{F B}=\p_{B}\Th_{F}+\p_{F}\Th_{B}~, 
		&&\Omega_{B F^*}=\Omega_{F^* B}=\p_B\Th_{F^*}+\p_{F^*}\Th_{B}~,\\
	&\Omega_{B^* F}=\Omega_{F B^*}=\p_{B^*}\Th_{F}+\p_{F}\Th_{B^*}~,
		&&\Omega_{B^* F^*}= \Omega_{F^* B^*}=\p_{B^*}\Th_{F^*}+\p_{F^*}\Th_{B^*}~, \\
	& \Omega_{F F}=2\p_{F}\Th_{F}~,\quad\ \Omega_{F^* F^*}=2\p_{F^*}\Th_{F^*}~,\quad
		&&\Omega_{F F^*}=\Omega_{F^* F} =\p_F\Th_{F^*}+\p_{F^*}\Th_{F}~,
\ea\ee
where all derivatives are left derivatives. We then obtain the symplectic matrix
\be\label{SM}
\Omega_{kl}=\frac{i}{4\m^2}\left( \begin{array}{cccc}
 0 & -A & \frac{B^*F^*}{E_B} & -F^*\\ A & 0 & - F & \frac{B F}{E_B}\\ \frac{B^*F^*}{E_B} & - F & -2B^* & 2E_B\\
 -F^* & \frac{B F}{E_B} & 2E_B& -2 B \end{array}\right)~,
 \quad \mbox{with} \quad A=\frac{4\m^2-F^*F}{2E}~.
\ee
This equations  generalize the symplectic form \eqref{2 Form} for the $\mbox{OSP}(1|2)$ coadjoint orbits.

According to \eqref{psu(1,1.1) algebra by PB} the matrix formed by the Poisson brackets $\Omega^{kl}=\{\xi^k,\,\xi^l\}$
reads
\be\label{PB SM}
\Omega^{kl}=\left( \begin{array}{cccc}
 0 & -2i E & 0 & -i F\\ 2i E & 0 & i F^* & 0\\ 0 & -i F^* & -2iB & -2i E\\
 i F & 0 & -2i E& -2i B^* \end{array}\right)~.
\ee
It is straightforward to check that this matrix inverts the symplectic matrix \eqref{SM},
demonstrating the consistency of the calculations.

Now we consider the massless superparticle on AdS$_2$. Here, as for the bosonic case (see \eqref{massless data}), one should analyze
the 1-form and the Noether charge given by
\be\label{massles superpartical data}
\Th=\langle T_+ \,\mbox{d}g_r \,g_r^{-1}\rangle~, \qquad R=g_r^{-1}\,T_+\,g_r~.
\ee
We use the representation $g_r=g_f g_b$, where $g_f=e^{\th_+ S_-+\th_- S_+}$,  as in \eqref{gf}, and
$g_b$ is parameterized similarly to the massless bosonic case $g_b=e^{2\a T_+}\,e^{\frac{\b}{2}\,T_1}\,e^{2\g T_-}$.
 The 1-form $\Th$ then splits into a sum of fermionic and bosonic differentials $\Th=\langle\,T_+\mbox{d}{g}_f \,g_f^{_{-1}}\,\rangle+\langle\,g_f^{_{-1}}\,T_+\,g_f\,\mbox{d}g_b\,g_b^{_{-1}}\,\rangle$.

 Using the expansion
 \be\label{gf=}
 g_f=I+\th_+ S_-+\th_- S_+ + \frac{\th_+\th_-}{2}(S_-S_+ - S_+S_-)
 \ee
 and the algebra \eqref{comm T1}, we find
 \be\label{f-orbit of T+}
 g_f^{_{-1}}\,T_+\,g_f=(1-2i\th_+\th_-)T_+ + \th_+S_+~,
 \ee
 \be\label{f-orbit of 1-form}\nonumber
 \mbox{d}g_f\,g_f^{_{-1}}=2i(\th_+\mbox{d}\th_+\,T_- -\th_-\mbox{d}\th_-\,T_+) -i(\th_+\mbox{d}\th_- + \th_-\mbox{d}\th_+)T_1
 +(1+i\th_+ \th_-)\left(\mbox{d}\th_+\,S_- + \mbox{d}\th_-\,S_+\right)~.
 \ee
 The calculation of the bosonic part is similar to \eqref{massless data 1}. Finally we obtain
 \be\label{massles superparticle data 1}
 \Th=i\th_+\mbox{d}\th_+ +e^\b(1-2i\th_+\th_-)\mbox{d}\g~,
 \ee
 \be\label{massles superparticle data 2}
 R=e^\b(1-2i\th_+\th_-)(T_+ - 2\g T_1-4\g^2T_-)+e^{\b/2}\,\th_+(S_+-2\g S_-)~.
 \ee
 Introducing the new bosonic variable by $e^{\tilde \b}=e^\b(1-2i\th_+\th_-)$
 we conclude that the dependence on $\th_-$ drops out, which demonstrates the $\k$-symmetry
 discussed above.

\subsection{Quantization}
Let us introduce the standard bosonic and fermionic creation-annihilation operators
$(b^{\dag}, b)$ and $(f^{\dag}, f )$, which satisfy the canonical commutation relations
$[b, b^{\dag}]=1$ and $\{f, f^{\dag}\}=1$.
The operators for the Noether charges are defined on the basis of the classical representation
\eqref{b,f to B,F} and the operator ordering freedom is fixed by
\be\ba\label{q supersymmetry generators}
&E=\m+b^\dag b+\frac{f^\dag f}{2}~,\\
&B=\sqrt{2\m+b^\dag b+f^\dag f}\,\,b~,
\qquad && \quad B^*=B^{_\dag}~,\\
&F=\sqrt{2\m+b^\dag b+f^\dag f}\,\,f+f^{\dag}\,b~,  && \quad F^*=F^{_\dag}~.
\ea\ee
Note that the classical expressions for $B$ and $F$ in \eqref{b,f to B,F} can also be written in
this form.
This form of the symmetry generators becomes helpful for calculating of commutation relations.

The operators \eqref{q supersymmetry generators} act in the Hilbert space spanned by the energy eigenvectors $|n,m\rangle$,
with $n\geq 0$ and $m=(0,1)$. The energy spectrum, therefore, is
\be\label{E spactrum}
E_{nm}=\m+n+\frac{m}{2}~.
\ee
The action of the operator $\sqrt{2\m+b^\dag b+f^\dag f}$ on the energy eigenstates is defined as in the Holstein-Primakoff representation by
\be\label{sqrt op}
\sqrt{2\m+b^\dag b+f^\dag f}\,|n,m\rangle=\sqrt{2\m+n+m}\,|n,m\rangle~.
\ee

It is straightforward to check that the operators \eqref{q supersymmetry generators} satisfy the commutation relations
of the $\mathfrak{osp}(1|2)$ algebra
\bea\nonumber
&&[E,B^\pm]=\pm \,B^\pm~,   \qquad  [B^-,B^+]=2E~,\\ \label{q superalgebra}
&&[E,F^\pm]=\pm\f{1}{2}\,F^\pm~, \quad \ [B^\pm, F^\pm]=0~, \quad \ [B^\pm, F^\mp]=\mp F^\pm ~, \\ \nonumber
&&\{F^+,F^-\}=2 E~, \qquad \{F^\pm,F^\pm\}=2\,B^\pm~,~~~~~~~~~
\eea
where we have defined $B^-=B$, $B^+ =B^*$, $F^-=F$ and $F^+ =F^*$.

The calculation of the quantum Casimir number from \eqref{q supersymmetry generators} yields
\be\label{Casimir qu}
C=E^2-\frac{1}{2}(B^-B^+ + B^+B^-)-\frac{1}{4}(F^+F^--F^-F^+)=\m(\m-1/2)~.
\ee

The massless case corresponds to $\mu=0$. As in the bosonic case the vacuum is invariant
under the action of all symmetry generators. Therefore, to construct an irreducible representation,
one has to remove the state $|0, 0\rangle$. One can show that the resulting representation is unitary equivalent to the representation \eqref{q supersymmetry generators} at $\m=1/2$.

\section{Conclusion}\label{conclusion}
In this article we have canonically quantized a massive AdS$_2$ superparticle
on the basis of the superisometry group OSP$(1|2)$, generalizing the
construction for the bosonic particle on AdS$_2$. Gauging an SO$(1,1)$ subgroup
of OSP$(1|2)$, we considered the action given by the square of
the left current projected onto the bosonic part of the coset.
	For the massive case, we represented the mass-shell phase space as a coadjoint orbit of a timelike element of $\mathfrak{osp}(1|2)$, giving a well defined symplectic structure and a realization of the OSP$(1|2)$ symmetry as the Poisson bracket algebra of the Noether charges.
	Our parametrization immediately yielded a description in terms of one bosonic and one fermionic oscillator and their canonical quantization led to a Holstein-Primakoff type realization of $\mathfrak{osp}(1|2)$.
	
	Repeating the analysis for the massless case, we observed the decoupling of one fermion, which is an explicit demonstration of $\kappa$-symmetry in our setting. As this leaves only one real fermionic field, quantization of this system appears inconsistent.

%
	There are a number of natural generalizations that would be of interest to explore in the future. One immediate open question is the quantization of the massless superparticle in AdS$_2$.
	The obstructions encountered due to the $\kappa$-symmetry suggests that one ought to consider the larger group SU$(1,1|1)$, gauging an SO$(1,1) \times \mbox{U}(1)$ subgroup, as considered in, for example, \cite{Adam:2007ws}.
One could  also consider an alternative $\mathbb{Z}_4$ grading for which an
SO$(1,1)$ subgroup of SU$(1,1|1)$ is gauged. The resulting model describes a superparticle on AdS$_2 \times \mbox{S}^1$ and has been subject of the works \cite{Galajinsky:2010zy, Galajinsky:2011xp, Orekhov:2014xra} relevant for the Kerr/CFT correspondence \cite{Guica:2008mu}. Taking
account of the $\kappa$-symmetries, the massless case of this model with fixed angular momentum on
the S$^1$ should be classically equivalent to the massive model considered in
this paper.

Furthermore, the theory considered in this paper can be understood as a
truncation of various supercoset models related to known critical superstring
backgrounds \cite{Zarembo:2010sg, Wulff:2014kja}. These include the supercoset
PSU$(1,1|2)/\mbox{SO}(1,1)\times \mbox{SO}(2)$, related to the AdS$_2 \times
\mbox{S}^2 (\times \mbox{T}^6)$ string background
\cite{Zhou:1999sm,Berkovits:1999zq}, and D$(2,1;\alpha)/\mbox{SO}(1,1)\times
\mbox{SO}(2) \times \mbox{SO}(2)$, related to AdS$_2 \times \mbox{S}^2 \times
\mbox{S}^2 (\times \mbox{T}^4)$.  Generalizing further to these cases may help
in understanding the connection to the full critical superstring theory. In
addition, in the first case it would be interesting to understand the relation
to the construction of \cite{Bellucci:2002va,Ivanov:2002tb}.

The extension to higher dimensional Anti de Sitter spaces is an important
next step. The case of AdS$_3$ could be a helpful stepping stone in this direction as in the minimal
case the isometry group of the superparticle action takes direct product
form, $\mbox{OSP}(1|2) \times \mbox{OSP}(1|2)$. Hence the results
of this paper relating to the supergroup $\mbox{OSP}(1|2)$ will be applicable
therefor.

Finally, let us conclude by recalling that one of the eventual aims of this
program is the application to AdS superstring theories, of interest in the
context of the AdS/CFT correspondence, and the quantization of strings on these
backgrounds from first principles.

\subsection*{Acknowledgements}
\noindent
We are grateful to Gleb Arutyunov, Harald Dorn, Sergey Frolov, Vladimir Mitev, Jan Plefka, Volker Schomerus, Arkady Tseytlin, Konstantin Zarembo and Alexander Zheltukhin for useful discussions.
M.H. thanks Nordita in Stockholm for kind hospitality.
G.J. thanks the Humboldt University of Berlin for kind hospitality.\\
M.H. is supported by the German Science Foundation (DFG) under the Collaborative Research Center (SFB) 676 Particles, Strings and the Early Universe.
B.H. is supported by the DFG through the Emmy Noether Program “Gauge Fields from Strings” and SFB 647 Space - Time - Matter, Analytic and Geometric Structures.
G.J. and L.M. 
have received funding from Rustaveli GNSF and G.J. from a DFG grant in the framework of the SFB 647. \\


\bibliographystyle{nb}
\bibliography{bibSpires}

\end{document}